\documentclass[superscriptaddress,aps,twocolumn]{revtex4-1}
\usepackage{bm}
\usepackage{float}
\usepackage{amssymb}
\usepackage{amsmath}
\usepackage{multirow}
\usepackage{lineno}
\usepackage{graphicx}
\usepackage{comment}
\usepackage{soul}
\usepackage[colorlinks=true,citecolor=blue,linkcolor=blue]{hyperref}
\usepackage[usenames]{color}
\usepackage{subfigure}

\begin{document}

\title{Helical Electron Beam Micro-Bunching by High-Order Modes in a Micro-Plasma Waveguide}

\author{Xingju Guo}
\affiliation{State Key Laboratory of Dark Matter Physics, Key Laboratory for Laser Plasma (Ministry of Education), Tsung-Dao Lee Institute $\&$ School of Physics and Astronomy, Shanghai Jiao Tong University, Shanghai 201210, China}
\author{Longqing Yi}
\thanks{lqyi@sjtu.edu.cn}
\affiliation{State Key Laboratory of Dark Matter Physics, Key Laboratory for Laser Plasma (Ministry of Education), Tsung-Dao Lee Institute $\&$ School of Physics and Astronomy, Shanghai Jiao Tong University, Shanghai 201210, China}
\affiliation{Collaborative Innovation Center of IFSA (CICIFSA), Shanghai Jiao Tong University, Shanghai 200240, China}

\date{\today}

\begin{abstract}
Electron acceleration by a high-power Laguerre-Gaussian pulse in a micro-plasma waveguide is investigated. When the incident laser travels in the waveguide, electrons on the wall are extracted into the vacuum core and accelerated by the longitudinal field of the waveguide mode. Using 3D particle-in-cell simulations, we demonstrate that high energy ($\sim 100$ MeV) electron beams with extremely high charge ($\sim 10$ nC), ultrashort duration ($\sim 30$ fs) and small divergence ($\sim 1^\circ$) can be produced by a 100-TW, few-Joule class laser system.
In particular, when the drive Laguerre-Gaussian pulse is circularly polarized, it excites high-order waveguide modes that exhibit helical longitudinal electric fields. The 3D profile of this accelerating field is imprinted into the high energy electron beam, leading to helical micro-bunching. This process can be controlled by the spin and orbital angular momentum of the drive pulse. This work paves the way to the generation of high-charge, relativistic electron beams with controlled helicity, which holds great potential for advances in fundamental science and a variety of applications.
\end{abstract}
\maketitle

Laser plasma accelerators can sustain large acceleration gradients up to hundreds of GVm$^{-1}$, orders of magnitude greater than that of a state-of-the-art radio-frequency accelerator\cite{hooker2013developments,esarey2009physics}. 
This makes them an attractive option for developing compact, table-top accelerators, which is key to improving accessibility to high-energy electron beams for a variety of applications in industry, medicine and fundamental research. Among these laser-based accelerating schemes, laser wakefield acceleration (LWFA) \cite{Tajima1979} can produce multi-GeV, quasi-monoenergetic electron beams with high quality \cite{mangles2004monoenergetic,geddes2004high,osterhoff2008generation,leemans2014multi}, making it possible for driving X-ray free electron lasers \cite{Wang2021,fuchs2009laser,huang2012compact,maier2012demonstration}. On the other hand, laser-solid interaction \cite{Liao2019} and self-modulated LWFA in near-critical-density plasma \cite{rosmej2020high,Krall1993} are often relied on to generate electron beams with high charge up to $\mu$C, board-band energy spectra and large divergence, such electron beams are of interest for developing positron sources \cite{Xu2016,Zhu2024,xu2020driving}, as well as producing high-power Terahertz pulses via coherent transition radiation \cite{Liao2019,happek1991observation}.

On the other hand, high-power laser interacting with a micro-plasma waveguide (MPW), provides a promising alternative for tabletop electron accelerator as the beam parameters fill in the gap between the aforementioned schemes.
The MPW is made of solid cladding (plastic or metal) and hollow core, with a diameter comparable to laser wavelength ($\sim 10~\mu m$). 
Previous works \cite{yi2016bright,Yi2016POP} have shown it can guide the incident high-power laser pulse as {\it waveguide modes} over long distances beyond the Rayleigh range, and electrons are accelerated directly by the longitudinal electric field of waveguide modes. 
The maximal acceleration energy is typically a few hundreds of MeVs, with board-band energy spectra and good directionality. 
A total charge of a few tens of nCs can be obtained, as these electron bunches are extracted from solid waveguide walls, such that they inherit the high-density nature\cite{naumova2004attosecond}.
These features have been demonstrated experimentally \cite{Snyder2019} and utilized to produce ultra-intense Terahertz pulses \cite{yi2019coherent}.

Moreover, since the electrons can only be accelerated in suitable phase of the waveguide modes, the electron beam shape are thus controlled by the laser pulse\cite{yi2016bright}. For instance, a linearly polarized laser can generate an attosecond electron bunch train spatially separated by half laser wavelength \cite{naumova2004attosecond,Thevenet2016,Yi2016SR}, and a circularly polarized driver induces helical micro-bunching of the accelerated beam \cite{yi2016bright,Zhang2021}.
However, to achieve full degree-of-freedom manipulation of the micro-bunching process, the capacity of exciting arbitrary high-order waveguide modes is necessary.
Although the plasma waveguide theory has been studied in previous works \cite{Shen1991,yi2016bright,yu2019generation}, to control the high-order mode excitation remains challenging. Because the drive laser energy naturally tends to be coupled into the fundamental mode with the smallest eigenvalue. To the best of our knowledge, this problem remains unsolved to date.

In this letter, we demonstrate that by employing a circularly polarized (CP) Laguerre-Gaussian (LG) pulse as the driver, only high-order modes are excited in the MPW, which can be used to induce controlled helical micro-bunching of electron beams, by varying the total angular momentum of the LG pulse. 
This approach possesses the capability of producing ultra-dense relativistic electron beams with customized helical micro-bunching, and a maximum energy of 100s MeVs, total charge above 20 nC, divergence $\sim2^\circ$, which is valuable for fundamental science \cite{Ivanov2022} and a variety of applications\cite{Hemsing2014,Hemsing2013,Hemsing2009,Duris2014}.\\

We first present our theory on high-order waveguide modes. 
When the drive pulse is relativistically-strong, a strong evanescent wave is present at the plasma-vacuum interface, which changes the boundary condition.
In particular, the spatial structure of the evanescent wave, which is typically helical for a CP driver, determines the waveguide modes that can propagate in a MPW. 
By considering the plasma response to the propagating electromagnetic wave, the eigenvalue equation can be obtained as\cite{Shen1991}
\begin{equation}
\begin{aligned}
	A(m)[B(m)-2C(m)]+\frac{\omega_p^2}{\omega_0^2}C(m)K(m) = 0,
\end{aligned}
\end{equation}
where 
\begin{equation}
	\begin{aligned}
		A(m)=J(m)+K(m),
	\end{aligned}
\end{equation}
\begin{equation}
	\begin{aligned}
		B(m)=A(m)+\frac{\omega_p^2}{\omega_0^2}\left[\frac{m}{y^2}-K(m)\right],
	\end{aligned}
\end{equation}
\begin{equation}
	\begin{aligned}
		C(m)=\frac{m}{x^2}+\frac{m}{y^2},
	\end{aligned}
\end{equation}
\begin{equation}
	\begin{aligned}
		J(m)=\frac{J_{m+1}(x)}{xJ_m(x)},~ K(m)=\frac{K_{m+1}(y)}{yK_m(x)}.
	\end{aligned}
\end{equation}
Here $\omega_0$ is the laser frequency, $\omega_p = \sqrt{4\pi n_ee^2/m_e}$ is the plasma frequency of the cladding, and $y = \sqrt{(cr_0/\omega_p)^2-x^2}$, with $c$, $e$, $m_e$, $n_e$ and $r_0$ denotes vacuum light speed, elementary charge, electron mass, plasma density and MPW radius, respectively. $J_m(x)$ is the Bessel function of the first kind, and $K_m(y)$ is the modified Bessel function of the second kind.

\begin{figure*}[t]
	\centering
	\includegraphics[width=0.95\textwidth]{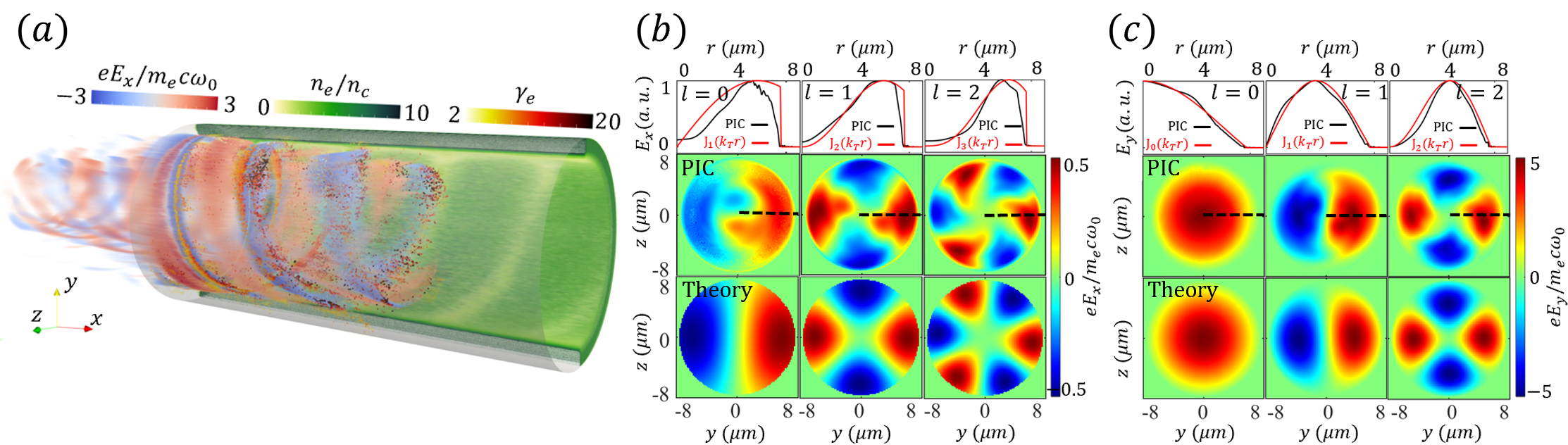}
	\caption{
		(a) Schematic sketch of the waveguide mode excitation, where the 3D color-coded acceleration field $E_x$ (blue-red), and surface electron density fluctuation (the evanescent wave, in white-green) are presented. The orange dots represents the injected electrons. The drive laser pulse is RCP with $l=1$.
		The distribution of (b) $E_x$ and (c) $E_y$ fields, where the first, second, and third columns corresponding to topological charges $l=0,1,2$ of the incident laser pulse. The second and third row show the results from PIC simulations and Eq.~(6-7), respectively. In the first row of (b-c), the black lines illustrate the radial distribution of the fields in each case along the dashed lines, and the red lines show the prediction of the theory.
	}
	\label{fig:1}
\end{figure*}

The eigenvalue equation Eq.~(1-5) was first derived in Ref.~\cite{Shen1991} to study the transfer of electromagnetic wave, and has previously been applied in the study of electron acceleration \cite{Yi2016SR} and X-ray emission \cite{yi2016bright,Yi2016POP}, but the implication on high-order waveguide modes and their application have never been discussed.
By solving Eqs.~(1-5), one obtains the roots $x_{|m|,n}$, where the subscript $|m|$ denotes the absolute value of the azimuthal mode number, and $n$ is the radial mode number, namely, $x_{|m|,n}$ is the $n$th root for a given $|m|$.

Notably, the value of $m$ can be negative in Eqs.~(1-5), since the left-hand side of Eq.~(1) is a even function for $m$, normally one does not distinguish the sign of $m$. However, in this study we found the sign of $m$ is important because it is related to the handedness of the helical micro-bunching and the excitation of high-order radial modes $n>1$ (discussed later). So we shall follow the conventional notation to use $|m|$ in the subscript of eigenvalue and waveguide modes, but present our main findings based on the actual value of $m$.

Since both electric and magnetic fields are presented in the propagating direction, the waveguide mode can be written as HE$_{|m|,n}$, with a transverse wavenumber of $k_T = x_{|m|,n}/r_0$. 
We write the electric field of the waveguide mode in Cartesian coordinates with $+x$ being the propagating direction.
\begin{equation}
	\begin{aligned}
		E_x=&\frac{k_T}{k_0}E_mJ_m(k_Tr)\\
		&\times\exp(ik_xx-i\omega_0t+i\sigma m\phi-i\frac{\pi}{2}),
	\end{aligned}
\end{equation}
\begin{equation}
	\begin{aligned}
	\mathbf{E}_\perp\approx&(\mathbf{e}_y+i\sigma\mathbf{e}_z)E_mJ_{m-1}(k_Tr)\\
	&\times\exp\left[ik_xx-i\omega_0t+i\sigma(m-1)\phi\right],
	\end{aligned}
\end{equation}
where $E_m$ is the amplitude of the transverse electric field of the mode, $r=\sqrt{y^2+z^2}$ is the radial coordinate, and $\phi$ is the azimuthal angle with respect to $y$ axis. The laser wavenumber in vacuum is $k_0 = c/\omega_0 = \sqrt{k_x^2+k_T^2}$, and $\sigma=+1$ and $-1$ corresponding to right-handed CP (RCP) and left-handed CP (LCP) of the drive laser pulse. Note that both $E_x$ and $\mathbf{E_\perp}$ have helical phase front, that depend on the spin $\sigma$ and azimuthal mode number $m$.

Because the generated waveguide modes must satisfy initial condition, the field structure at the entrance of MPW is determined by the incident laser field $\mathbf{E}_l$. In particular, the terms depending on $\phi$ in Eq.~(7) must be the same as in $\mathbf{E}_l$. This explains why previous studies \cite{yi2016bright, Yi2016POP} that typically employ a planar wave as the driver only consider the modes with $m=1$.

In order to excite high-order modes ($m\neq1$), we consider a circularly polarized LG pulse as the driver\cite{allen1992orbital} : 
\begin{equation}
	\begin{aligned}
		\mathbf{E}_l=(\mathbf{e}_y+i\sigma\mathbf{e}_z)&E_0(\frac{\sqrt{2}r}{w_0})^l\exp(\frac{-r^2}{w_0^2})\sin^2(\frac{\pi t}{2\tau_0})\\
		&\times\exp(ik_0x-i\omega_0t+il\phi),
	\end{aligned}
\end{equation}
where $E_0$, $w_0$, $\tau_0$, and $l$ represent the amplitude, focal spot size, duration, and the topological charge of the driver, respectively. By equating the $\phi$-dependent terms in Eq.~(7-8), one can easily obtain the azimuthal mode number that can satisfy this initial condition is 
\begin{equation}
	\begin{aligned}
        m=(l+\sigma)\sigma.
	\end{aligned}
\end{equation}

In this work, we  focus on the the circularly polarized LG drivers, as the linearly polarized pulses pulse can not produce helical micro-bunching for the accelerated electron beams. Also, from a theoretical point of view, a linear polarized LG pulse can be decomposed into a RCP and a LCP pulse, which lead to double mode excitation (see Supplementary Materials) with different $m$ according to Eq.~(9).\\


In the following, we demonstrated this high-order mode waveguide theory with 3D PIC simulations as illustrated in Fig.~1, where we show RCP ($\sigma=1$) LG beams with three different topological charges $l = 0,1,2$ entering a MPW with radius $r_0 = 8 {\rm\mu m}$. The normalized laser amplitude adopted in the simulations is $a_0 = eE_0/m_e\omega_0c = 15$, corresponding to an intensity of $I_0 = 6.2\times 10^{20}$ W/cm$^2$, the laser wavelength is $\lambda_0 = 1 {\rm\mu m}$, and spot size $w_0 = 0.8r_0 =6.4 {\rm\mu m}$, pulse duration $\tau_0 = 42$ fs (FHWM is 30 fs). The laser enters a pre-ionized MPW [assumed plastic (CH)] at $x_0 = 5 {\rm\mu m}$, and propagates to the $+x$ direction. The electron density is $n_0 = 20n_c$ in the MPW cladding, where $n_c = m_e\omega_0^2/4\pi e^2\approx1.1\times10^{21}$ cm$^{-3}$ is the critical density. A density gradient is present at the inner surface, $n_e(r) = n_0\exp[(r-r_0)/h]$ for $r_0-2 {\rm\mu m}<r<r_0$, with $h = 0.2\mu m$ the scale length.
The simulation is performed with a fully kinetic PIC code {\sc EPOCH} \cite{arber2015contemporary}. The size of the simulation box is $L_x\times Ly\times Lz = 32 \times 12 \times 12~{\rm\mu m}^3$, which is sampled by $1280 \times 240 \times 240$ cells, with $4$, $2$, and $2$ macro-particles per cell for electrons, $C^{6+}$ and protons, respectively.

As a primary example, Fig.~1(a) shows the color-coded longitudinal electric field and the electron density fluctuation (evanescent wave) near the entrance of the MPW.
One can see an increase of $E_x$ field strength as the laser enters waveguide mode, which is important for the electron acceleration. Moreover, the laser longitudinal field exhibits double-helix structure  $E_x\sim \exp[i(l+\sigma)\phi]$, as predicted by Eq.~(6) and (9). The $E_x$ field structure matches the evanescent wave on the inner surface, suggesting boundary condition is satisfied.

Similarly, for topological charges $l = 0$ and $2$ (both with $\sigma = 1$), we obtain $m = 1$ and $3$ as shown in Fig.~1(b), agreeing very well with the theory.
Notably, unlike in the conventional metallic waveguide, the $E_x$ field is nonzero at the MPW boundary $r=r_0$ [Fig.~1(b)]. Instead, it drops abruptly to zero within a small distance, which is an indication of a strong evanescent wave.

\begin{figure*}[t]
	\centering
	\includegraphics[width=0.9\textwidth]{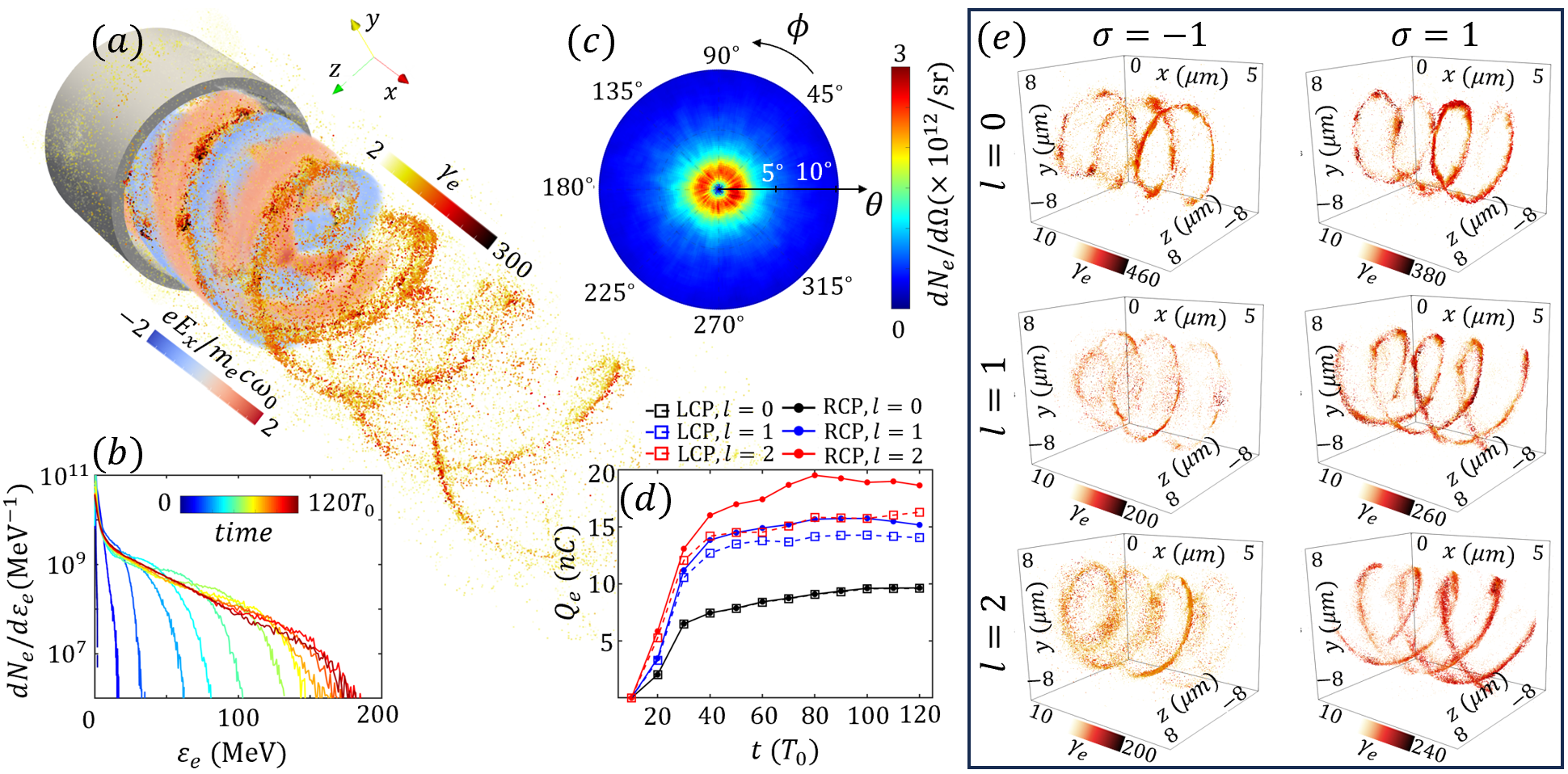}
	\caption{
		(a) The 3D structure of the accelerated electrons (color-code with energy) and accelerating field $E_x$ at $t = 120T_0$, when the electron energy reaches maximum. (b) The temporal evolution of the accelerated electron spectrum and  (c) angular distribution at $t = 120T_0$. (d) The total charge and (e) different helical structures of high-energy ($\gamma_e>10$) electron beams accelerated by drive pulses with various polarization and topological charges.
	}
	\label{fig:2}
\end{figure*}

We then proceed to check the radial profile of the electromagnetic fields in the MPW. Equations~(6-7) suggest the longitudinal and transverse electric fields of azimuthal mode number $m$ are proportional to $J_m(k_Tr)$ and $J_{m-1}(k_Tr)$, respectively. This relation is tested in the first row of Fig.~1(b-c). Moreover, the second and third rows present a comparison of the transverse electric field profiles between the PIC simulation data and that calculated from Eqs.~(6-7). Apparently, the theory matches the data very well.\\

When the drive laser pulse is sufficiently strong, the electrons in the plasma cladding can be pulled out, forming over-dense electron bunches that can be accelerated subsequently by the longitudinal electric field\cite{Gong2019}.
In Fig.~1(a), the injected electrons ($\gamma>10$) at the entrance of MPW are represented by the orange-white dots.
These electrons are rapidly boosted to relativistic energies, such that they keep up with the accelerating phase ($E_x<0$) throughout the acceleration, until dephasing takes place.
As a result, the 3D shape of the accelerating field $E_x$ is imprinted into the spatial structure of the high-energy electron beam, leading to helical micro-bunching. This is illustrated in Fig.~2(a), where the electron beam structure after $120-\mu m$ propagation in the MPW is presented, the simulation parameters are the same as shown in Fig.~1(a).

One can see that a double-helix beam structure arises, matching the $E_x$ field shape of the HE$_{2,1}$ mode, with a maximum energy reaching $150$ MeV.
In general, the micro-bunching of the electron beam is expected to have a helical structure as indicated by Eq.~(6), which can be controlled by the total angular momentum ($l+\sigma$) of the drive pulse.
	
The temporal evolution of the electron energy spectrum is represented in Fig.~2(b), which shows the electrons continuously gain energy as they propagate inside the MPW for approximately $120~\mu m$. The acceleration terminates due to dephasing between the electron and the superluminal accelerating field. This can be understand by comparing Fig.~2(a) and Fig.~1(a), where the electrons are injected uniformly into the accelerating bracket ($E_x < 0$) at the entrance of MPW [Fig.~1(a)], but they are shifting backwards with respect to the accelerating field as they propagate. By the time of $\sim 120~T_0$, most of them have fallen into the nodes of accelerating field ($E_x\approx 0$) [Fig.~2(a)]. Therefore, the dephasing length is $L_d \approx 4\pi^2r_0^2/x_{m,n}^2\lambda_0$, which corresponding to the relative shift between accelerating electrons and the waveguide mode reaches $\lambda_0/2$ \cite{Yi2016POP}. 

Moreover, Fig.~2(b) suggests that despite the electron energy spectrum being broadband in general, the electron charge is very high ($\sim 10$ pC/MeV at $100$ MeV). This makes it possible to obtain a mono-energetic bunch by energy filtering \cite{DeChant2025}.
The angular distribution of the accelerated electron beam is illustrated in Fig.~2(c). Obviously, the electrons form a doughnut shape with a singularity at the MPW axis, which is a signature of helical beam. The beam divergence is around $\theta\sim2^\circ$, and the electron number per solid angle reaches $10^{12}~\rm{sr}^{-1}$.

Such a high beam charge sterns from the fact that electrons are extracted out from the solid-density plasma cladding of the MPW. Figure~2(d) indicates that the accelerated electron charge are typically $\sim10s$ nC for a 100-TW, few-Joule class laser system. The beam charge increases with the absolute value of topological charge $|l|$ of the driver. 
This is because the electromagnetic intensity of LG drivers forms a ring, where the radius increases with $|l|$. Therefore, the interaction between the laser and the plasma cladding is increasingly strong as $|l|$ grows, leading to an enhanced electron injection. In addition, the polarization also play a minor role, this is due to high-order radial modes, which will be discussed later.


In Fig.~2(e), a variety of helical micro-bunching electron structures are displayed, which are produced in our PIC simulations with different combinations of $l$ and $\sigma$. For all the cases presented here, the azimuthal structure of the accelerating field leads to helical micro-bunching $\sim\exp[i(l+\sigma)\phi]$, in consistent with our theory.\\

\begin{figure}[t]
	\centering
	\includegraphics[width=8.5cm]{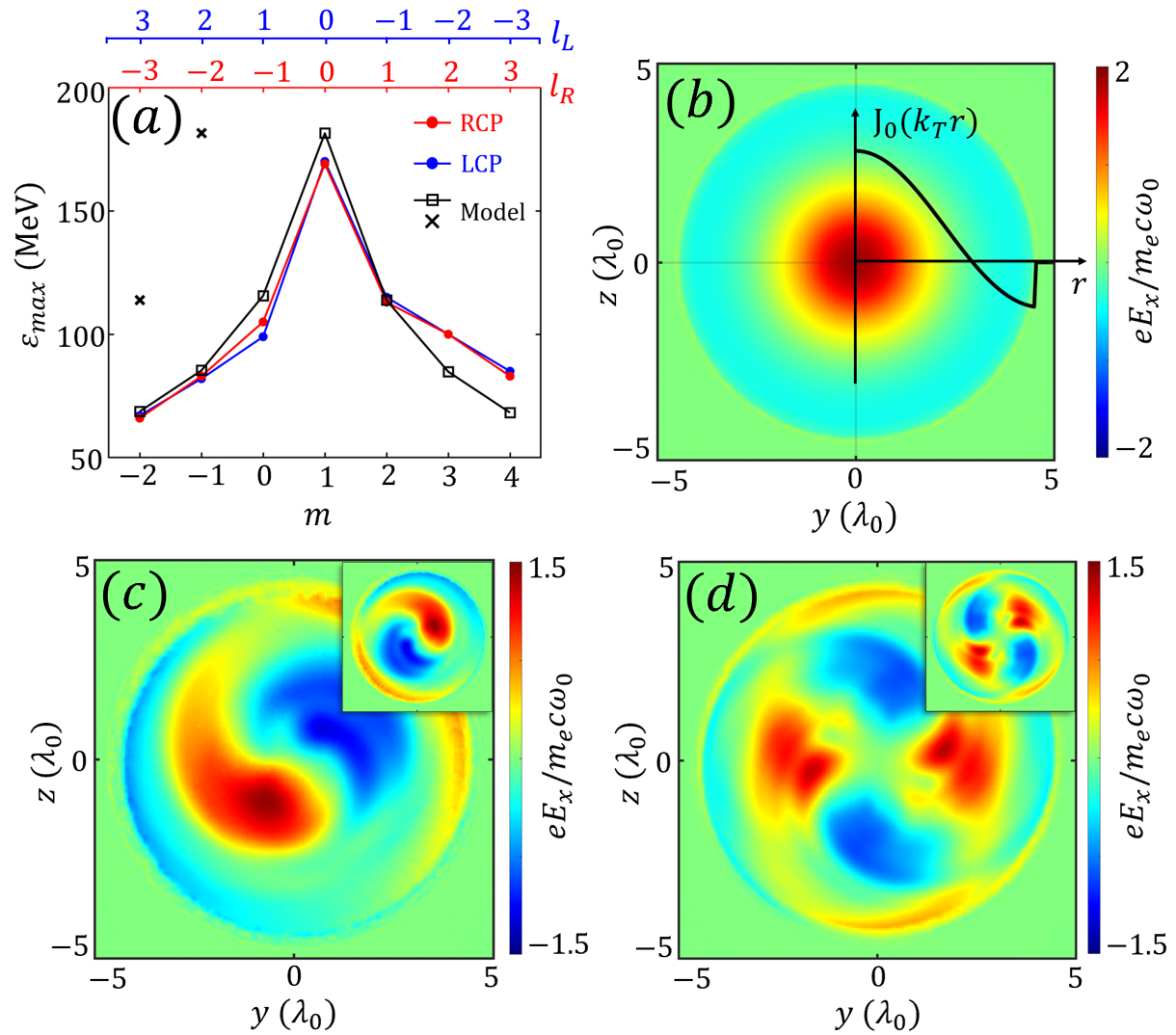}
	\caption{(a) The maximum energy plotted against azimuthal mode number $m$ for various polarization and topological charges. The red and blue color represents PIC results for RCP and LCP drivers, and the black color shows the prediction of our model. Note that for $m = -1$ and $-2$, the square markers shows the theoretical prediction of the second radial mode (with eigenvalue $x_{1,2}$ and $x_{2,2}$), and cross markers indicate the first radial mode ($x_{1,1}$ $\&$ $x_{2,1}$). (b), (c), and (d) show the acceleration field $E_x$ distribution with mode number $m = 0$, $-1$, and $-2$ for RCP drivers, respectively. The black line in (b) shows the prediction of our theory, and the insets in (c-d) shows the LCP case for the same $m$.}
	\label{fig:3}
\end{figure}

Finally, we discuss the maximum acceleration energy and the excitation of high-order radial mode with $n>1$ in the MPW.
The acceleration energy is $\varepsilon_{max}\approx (k_T/k_0)E_0L_d\propto x_{|m|,n}^{-1}a_0r_0$. Previous works \cite{Yi2016POP} have demonstrated the linear correlation of $\varepsilon_{max}$ with $a_0$ and $r_0$, here we will focus on the effects of high-order modes, in particular the eigenvalue $x_{|m|,n}$.
We thus scanned $l = -3$ to $3$ for $\sigma = \pm1$ in our PIC simulations, with the results represented by the red and blue curves in Fig.~3(a).

As one can see, for both RCP and LCP drivers, the maximum electron energy are achieved when $l = 0$, or equivalently $m = 1$. This is expected for $m\geq 0$ as Eq.~(1) suggests $x_{|m|,1}$ reaches minimum when $m=1$, namely $x_{1,1} \approx 2.4$ for plasma cladding density $n_e\sim 1n_c$. We note the variation of plasma density in the range $[1, 10]n_c$ has little effects on the eigenvalue, as discussed in the Supplementary Materials. Since the laser pulse is grazing incident on the plasma wall in the MPW, it is unlikely to penetrate deeply into the high-density region with $n_e > 10 n_c$.

When $m = 0$, Eq.~(6) suggests $E_x\propto J_0(x_{0,1}r/r_0)$ for the HE$_{0,1}$ mode, which decreases to zero faster than the HE$_{1,1}$ mode [$E_x\propto J_1(x_{1,1}r/r_0)$] with increasing $r$. Therefore, with an eigenvalue $x_{0,1}\approx 3.7 > x_{1,1}$, the accelerating field $E_x$ for the HE$_{0,1}$ mode reverse sign as $r$ increases from zero to the $r_0$, this is observed in the PIC simulations as shown in Fig.~3(b). Consequently, it produces a train of ring-shaped the electron bunches, as indicated by the case with $l=1$ and $\sigma = -1$ in Fig.~2(e).

When $m<0$, the PIC simulation data presented in Fig.~3(a) are counter-intuitive. Since the left hand side of Eq.~(1) is an even function for $m$, one would expect the maximum acceleration energies are similar for $m = \pm1$ and $\pm2$. This is indicated by the black crosses in Fig.~3(a), which are calculated from eigenvalue of the lowest radial mode $x_{1,1}$ and $x_{2,1}$, respectively. However, Fig.~3(c-d) show the dominate mode generated for $m=-1$ and $-2$ are both the second radial modes (HE$_{1,2}$ and HE$_{2,2}$), the signature being a reverse sign of the $E_x$ field as $r$ increases [see Fig.~1(b) for a comparison with HE$_{1,1}$ and HE$_{2,1}$ modes]. By employing the eigenvalues $x_{1,2}\approx 5$ and $x_{2,2}\approx 6.3$ in our theory, the resulted trend agrees with that observed in simulations.

The reason for the excitation of high-order radial modes is beyond the scope of the current work. However, we note this implies the phase velocities of RCP and LCP incident LG pulses with the same topological charge are almost the same. Therefore a linearly polarized LG laser (superposition of RCP and LCP pulses) traveling through a plasma waveguide does not change its polarization angle, which can serve as an experimental observable to be tested in future.\\

In conclusion, we have shown circularly polarized LG pulses can generate high-order azimuthal modes in a plasma waveguide. Overdense electron bunches are pulled out from the wall and accelerated by the longitudinal electric field of the waveguide modes. Therefore, the 3D helical structure of the high-order modes are imprinted to the high-energy electron beam, leading to helical micro-bunching of the beams. A high-order waveguide mode theory is derived in this work, which shows the azimuthal mode number is $m = (l+\sigma)\sigma$, and the helicity of the electron bunch controlled by the total angular momentum of the driver ($l+\sigma$). These results are verified in 3D PIC simulations, which paves the way towards future tabletop helical electron beam accelerator.\\

\begin{acknowledgments}
	This work is supported by the National Key R$\&$D Program of China (No. 2021YFA1601700), and the National Natural Science Foundation of China (No. 12475246).
\end{acknowledgments}




\bibliography{reference}



\end{document}